\documentclass[11pt]{article}
\usepackage[dvipdf]{graphics}
\def\physicaa#1#2#3{#3, { Physica A} #1, #2.}
\def\molsim#1#2#3{#3, { Mol. Sim.} #1, #2.}

\def\prl#1#2#3{{#3,  Phys. Rev. Lett.} #1, #2.}

\def\pre#1#2#3{#3, {Phys. Rev. E.} #1, #2.}

\def\jcp#1#2#3{#3, { J. Chem. Phys.} #1, #2.}
\def\jpc#1#2#3{#3, { J. Phys. Chem}  #1, #2.}
\def\jpcb#1#2#3{#3, { J. Phys. Chem. B} #1, #2.}

\def\cpl#1#2#3{#3, { Chem. Phys. Lett.}  #1, #2.}

\def\nature#1#2#3{#3,  { Nature}  #1, #2.}

\def\AA{A$^\circ$}
\def\onebyf{$1/f^\alpha$}

\def\avg#1{\langle #1\rangle}

\def\be{\begin{equation}}
\def\ee{\end{equation}}

\begin{document}
\baselineskip 22pt 

\begin{center}

{\bf\Large  Effect of the Berendsen thermostat on 
dynamical properties of water}
\vskip 0.5truein

{ Anirban Mudi and Charusita  Chakravarty$^{*}$}\\
Department of Chemistry, Indian Institute of Technology-Delhi,
New Delhi: 110016, India.\\
\end{center}

\vfill
{* Author for correspondence (Tel: (+) 91 11 26591510; Fax: (+) 91 11 2651 6459;
E-mail: {\tt charus@chemistry.iitd.ernet.in)}\hfill}

\newpage

\begin{center}
{\bf Abstract}\\
\qquad\\
\end{center}
The effect of the Berendsen thermostat on the dynamical properties
of bulk SPC/E water is tested by generating power spectra associated
with fluctuations in various observables.  The Berendsen thermostat is found
to be very effective in preserving temporal correlations in fluctuations
of tagged particle quantities over a very wide range of frequencies.
Even correlations in fluctuations of global properties, such as
the total potential energy, are well-preserved for time periods
shorter than the thermostat time constant. Deviations in dynamical
behaviour from the microcanonical limit do not, however, always decrease 
smoothly with increasing values of the thermostat time constant but
may be somewhat larger for some intermediate values of $\tau_B$, specially
in the supercooled regime, which are similar to time scales for
slow relaxation processes in bulk water.

\newpage

The ideal ensemble to extract dynamical information from a molecular dynamics
simulation is the microcanonical (NVE) ensemble \cite{at86,fs02}. 
Since the total energy
is conserved in this ensemble, the Newtonian equations of motion can
be assumed to represent the natural evolution of the system, subject
to the accuracy of using classical mechanics to describe the atomic
dynamics. The constant energy (E) and volume (V) conditions do not,
however, correspond to the most common 
experimental conditions and therefore
it is often desirable to implement MD simulations in the canonical (NVT)
 and isothermal-isobaric (NPT) ensembles. An additional reason for 
preferring the NVT  ensemble is that when long run lengths are required, 
as in the case of studies of  slow dynamics in liquids or glasses, 
there may be significant energy drift in the NVE ensemble.

Two types of approaches
have been developed to adapt MD simulations to the canonical ensemble:
(i) extended Lagrangian methods, such as Nose-Hoover thermostats 
and (ii) resampling or
 rescaling  of velocities, as in the case of the Andersen and 
Berendsen thermostats.
The extended Lagrangian methods will generate the true canonical
distribution of velocities. 
 The rescaling approaches will only ensure 
that the average kinetic energy of the system corresponds to  the expected value
at the desired temperature but have the advantage that they can
be combined very simply with the Verlet algorithm.
In both approaches,  the degree of perturbation
of the real time evolution of the system can be
adjusted by manipulating various thermostat parameters.  

The Berendsen thermostat represents a proportional scaling of the velocities
per time step in the algorithm with the scaling factor being given by
\be
\lambda =1 +\frac{\Delta t}{\tau_B}\Biggl( \frac{T_0}{T}-1\Biggr)
\ee
where $\Delta t$ is the time step, $\tau_B$ is the time constant of the
Berendsen thermostat , $T_0$ is the desired temperature and $T$ is the
instantaneous temperature \cite{bpvdh}.The velocity rescaling can be 
incorporated easily into the Verlet leapfrog algorithm. 
By varying the thermostat time constant, $\tau_B$, one
can, in effect, increase or decrease the degree of coupling 
to an external bath. The limit, $\tau_B\rightarrow\infty$, represents
the microcanonical ensemble. 
The original paper of Berendsen et al
applied this thermostatting procedure to simulations of 216 water molecules
bound by the rigid SPC potential. Static structural averages were
shown to be unaffected for values of $\tau_B$ as small as 0.01~ps.
Fluctutations in global properties were strongly  affected
when $\tau_B$ was less than 0.1ps implying that analysis of fluctuations
cannot be used to determine observable properties. Single-particle
properties, including dynamical quantities such as the diffusivity and
orientational correlation times, were found to  agree, within
the limits of statistical error,  with the results
derived from the microcanonical ensemble for $\tau_B > 0.1$ ps.
While the Berendsen thermostat is widely used, we have come across
only a limited set of subsequent studies in the literature which inspect
in detail  the nature of the associated trajectories
\cite{tm00,dta02,lb94}. 

In this note, we perform some  detailed tests of the effect of the
Berendsen thermostat on the dynamical properties of water. The motivation
for these tests comes from prior work  on molecular
dynamics simulations of bulk water where 
power spectra associated with a number of dynamical quantitites were
shown to possess a \onebyf\ frequency regime, indicative of
multiple time scale behaviour due to hydrogen bond network re-organisations
\cite{nrc95,sor92,io95,ss96,mrc03}.
The  power spectrum is defined as
\be
S(f) = {\left| \int_{t_{min}}^{t_{max}} \bigl(A(t)-\avg{A}\bigr)
e^{2{\pi}ift}{ dt}\right|}^2.
\ee
where  $A(t)$ is any time-dependent 
 mechanical quantity  and $\avg A$ is the corresponding average over the
system trajectory. Most recently we have examined power spectra
generated from fluctuations in tagged particle 
 quantities  sensitive to the local molecular environment, such as
the tagged particle potential and kinetic energies, over a wide
temperature range \cite{mrc03}.
Fluctuations in the tagged particle potential energies, but not
in the kinetic energies,  were shown to
give rise to \onebyf\ noise, indicative of a structural origin for
the multiple time-scale behaviour; moreover, the exponents and
frequency range of \onebyf\ behaviour were shown to be temperature-dependent. 
This suggests that the \onebyf\ behaviour of 
power spectra may provide a simple and  convenient
way of monitoring changes in the hydrogen bond network
dynamics  with temperature or density.  Since many of the
anomalous features of water, originating from its
unusual network structure, are more marked in the supercooled regime, it
is of interest to study  the \onebyf\ behaviour at low temperatures.
Molecular dynamics simulations of supercooled water typically require 
long run lengths of the order of 2ns or more if reliable estimates of
dynamical quantities, such as the diffusivity, are to be obtained. The Berendsen
thermostat is frequently used to generate the trajectory for such long
runs to avoid the energy drift associated with microcanonical runs and
facilitate comparison with experimental data
 \cite{sss99,hpss,nssb,nsbs,sslss}. 
To check the effect
of the Berendsen thermostat on  temporal correlations
associated with fluctuations in tagged particle quantities, we
have carried out a set of tests on a range of state points for
SPC/E water. 

The potential energy surface of water was represented by the  rigid,
nonpolarizable, three-site  SPC/E model \cite{bgs87}. 
The molecular dynamics (MD) simulations were  performed using
the  DL\_POLY software package \cite{fs02,syr01,dlp}. 
A cubic simulation cell containing 256 SPC/E water 
molecules was used. Electrostatic interactions were evaluated using 
the Ewald sum approach. The MD trajectory was propagated in the
microcanonical (NVE) ensemble using the Verlet leapfrog algorithm in 
conjunction with the SHAKE algorithm to implement bond constraints
\cite{at86}. MD trajectories were also generated 
in the NVT ensemble  using the Berendsen thermostat with a range of time
constants. In keeping with earlier work, a time step of 
1 fs was used for all the simulations and production
run lengths were kept at 2ns.  The coupling constant for the
Berendsen thermostat was assigned values of 1, 10, 25, 50 and 200ps.
Table 1 summarises our results from
from the NVT and NVE simulations at the following temperatures along
the 1 g cm$^{-3}$ isochore: 229K, 260K  and 295K.
The melting point of SPC/E water at 1 atm pressure is known to be below 260~K 
\cite{gsh02}.
 The pressure, $P$, and the configurational
energy, $U$, correspond to simple thermodynamic averages and
the associated errors were estimated by block averaging \cite{fp89}.
The large error bars on the pressure are typical of systems
with electrostatic interactions; within statistical error, our
results agree well with those in the literature.

We now consider the power spectra generated from fluctuations
in the tagged particle potential energies at the different state points
(see Figure 1). 
The tagged particle potential energy, $u(t)$, is defined as  the
interaction energy of an individual molecule with all the other 
molecules in the system and the corresponding power spectrum is
denoted by $S_u(f)$. Since the potential energy surface is assumed to be 
pair-additive, the total potential
energy, $U(t) =0.5\sum_i u_i(t)$ where the sum extends over all molecules. 
Tagged particle potential energies
of 32 molecules in each simulation were sampled at
intervals of 10fs and Fourier transformed to generate the power spectra
using standard FFT routines \cite{numrec}.
When generating power spectra,  a square time window of 328 ps width
 and a sampling interval of 10 fs was used.  This choice of
sampling interval corresponds  to a Nyquist
frequency of 1666 cm$^{-1}$ and sets an upper bound on the
frequency range over which we can obtain power spectra. 
The values of $\tau_B$  provide  the lower  limit on the 
frequency range over which we can obtain reliable power spectra;
thus, $\tau_B=1$ps corresponds to a lower frequency limit of 33 cm$^{-1}$. 
The normalisation convention
was chosen such that the integrated area under the $S(f)$ curve 
equalled the mean square amplitude of the time signal. Statistical
noise in the power spectra was reduced by averaging over overlapping
time signal windows as well as over individual tagged particle spectra;
typically 11 windows were used to cover the 2ns run length.
In a given frequency interval showing \onebyf\ behaviour, 
linear least squares fitting of $\ln S(f)$ was
done to obtain the $\alpha$ values with an estimated standard error
of less than 3\%. Additional computational details regarding calculation
of power spectra are given in ref.\cite{mrc03}.

Figure 1(a) shows the $S_u(f)$ spectra at 230K obtained from the NVE
run as well as from NVT runs with different values of $\tau_B$ are shown.
At this temperature, the $S_u(f)$ curve shows three distinct features:
(i) a broad peak, originating from librational modes, centred at 
approximately 600 cm$^{-1}$;
(ii) a \onebyf\ regime between 60  and 298 cm$^{-1}$ and (iii)
a second, low-frequency $1/f^{\alpha'}$ region between 1 and 40 cm$^{-1}$
such that $\alpha\neq\alpha'$.
On a logarithmic scale, the  power spectra obtained from the different
simulations are indistinguishable.
Careful inspection of the $S_u(f)$ curves, however, shows that the noise in
the results obtained with $\tau_B =25$ps is somewhat larger than in all the
other cases. To quantify the effect on the $S_u(f)$ curves, we have tabulated
the $\alpha$ and $\alpha '$ values in Table I. 
The low frequency exponent $\alpha'$ from all
the runs agrees to within 3\% except at $\tau_B =25$ps. The spread
in the values of the high-frequency exponent $\alpha$ is somewhat larger with
again a maximum deviation at $\tau_B=25$ps. While the deviation at 
$\tau_B=25$ps is not large, it is interesting that it should be greater than the
deviation seen for the smallest $\tau_B$ value of 1ps. 
The results for 260K,
shown in Figure 1(b) and Table I,
 are very similar in that variations with $\tau_B$ values are very
small and within the estimated statistical error. The $\alpha$ and $\alpha'$
values are, however, very similar, indicating that it may be better to
characterise the region from 1 to 300 cm$^{-1}$ as a single \onebyf\
region. The $S_u(f)$ curve at 295K clearly shows only one \onebyf\ region
which merges into the librational band with a crossover to white noise
behaviour below 2 cm$^{-1}$. 

In addition to the results shown for the tagged particle potential
energy, we have checked the convergence, with respect to
$\tau_B$,  of power spectra associated with a number of 
other tagged particle quantities, such as the local kinetic energies
and order parameters, and found similar trends to those discussed above.
From our results, it is evident that the Berendsen thermostat preserves
the temporal correlations in tagged particle quantities remarkably well
over a wide frequency range, even for Berendsen thermostat coupling constants
as small as 1 ps but  the small discrepancy seen at 230K for 
$\tau_B=25$ps  does remain.
Figure 2 shows the mean square displacement as a
function of time at this state point, as an additional check, from 
the different simulations. On a logarithmic scale, the curves are 
indistinguishable but it can be seen that the diffusivity estimates
obtained at $\tau_B=25$ps would be slightly different than for 
all the other cases.

To understand the discrepancy seen for $\tau_B =25$ps at 230K, we have
performed simulations for three additional values of $\tau_B$ at 230K
and 1 g cm$^{-3}$; all the tagged particle power spectra are shown
in Figure 3. It can be seen that the $\tau_B$ values of 25ps and 20ps
result in  significantly greater noise and change in shape of the
$S_u(f)$ curves in the 100 to
400 cm$^{-1}$ regime. These values of $\tau_B$ must therefore
correspond to some physically significant dynamical time scales of bulk
water. The frequencies of the O-O stretching and O-O-O bending modes 
are at 200 cm$^{-1}$ (0.165 ps)  and 50 cm$^{-1}$ (0.66ps) respectively
\cite{ph98}.
Clearly all the $\tau_B$ values chosen by us  lie above the 
time scales associated with these intermolecular vibrations.
In reviewing the literature on the dynamics of supercooled SPC/E water,
we found, however, that the time at which the non-Gaussianity
parameter is a maximum is 20.6 ps at 224K and 15.5ps at 238K \cite{sgtc}.
The time at which the non-Gaussianity parameter peaks is 
an indication of the time scale on which  dynamical correlations
are most pronounced.  This suggests that anomalous noise effect seen for 
$\tau_B$ values of 20 and 25ps is due to a coincidence of the time scales 
for the thermostat and for correlated motion in supercooled water.
At higher temperatures, such long-lived correlations are much less
pronounced and therefore variations in the power
spectra with $\tau_B$ are not as noticeable.

Figure 4 shows the power spectra obtained from fluctuations in the 
total potential energy of the system from NVE and NVT simulations.
As expected, these power spectra are much more noisy than those 
obtained from tagged particle potential energies. Despite the
fact that fluctuations in global quantities are expected to be
much more subject to distortion on thermostatting than single-particle
quantities, the power spectra obtained from the NVT ensemble simulations
with $\tau_B=200$ps agree well with the NVE results for a frequency
range of 0.2 to 1000 cm$^{-1}$. The results
for  $\tau_B=1$ps, do, however, show large deviations
in the low frequency regime below 10 cm$^{-1}$. 

This study indicates that NVT ensemble simulations
using the Berendsen thermostat are successfully able to
reproduce the temporal correlations in tagged particle quantities over a 
very wide frequency range provided sufficiently large values of the
Berendsen time constant are used. Even the power spectra associated with
global quantities, such as the total potential energy, are well reproduced
for frequencies greater than $1/\tau_B$. It is, however, worthwhile to
check the dynamical behaviour for a range of thermostat time constants since 
some values may show slightly larger deviations from microcanonical 
behaviour than others. This effect is more pronounced on supercooling
and is noticeable for values of the Berendsen thermostat time constant  which
are similar to the time at which the non-Gaussianity parameter is maximum,
suggesting  that the  anomalous  values of $\tau_B$ fall in the range of time
scales associated with slow structural relaxations. Though the Berendsen
thermostat is widely used, we are not aware of any previous study which
has demonstrated this effect. We expect that analogous deviations
from microcanonical behaviour will be seen with other thermostats 
when slow relaxation processes are present.

{\bf Acknowledgements} Financial support for this work has
been provided by the Department of Science and Technology (SP/S1/H-16/2000).
AM would like to thank the Council for Scientific and Industrial Research,
New Delhi for the award of a Junior Research Fellowship.

\newpage

\begin{table}
\caption{Summary of simulation results from microcanonical(NVE) and 
canonical(NVT) ensemble simulations
of SPC/E water at state points along  the 1 g cm$^{-3}$ isochore.
Each run is 2ns long. The error bars on the temperature ($T$) apply 
in the case of the microcanonical (NVE) ensemble run.
The exponent $\alpha$ is obtained by fitting $S_u(f)$ to a \onebyf\
form over the frequency ranges 60-298 cm$^{-1}$ (230K), 60-298 cm$^{-1}$ (260K)
and 4-200 (295K). The exponent $\alpha'$ is obtained by fitting to
a \onebyf\ form in the frequency range 1-40cm$^{-1}$ at 230K and 260K.}
\bigskip

\begin{tabular}{lllclll}
\hline\hline

\smallskip \\
T  &  $\tau_{B}$   & P  &    U    &  $\alpha$    & ${\alpha}'$\\
(K) & (ps)         & (MPa) & (kJ/mol) & \        & \           \\
\smallskip \\
\hline
229.5$\pm$0.7 & $\infty$ & -22$\pm$4 & -51.3$\pm$0.2 &  1.37     & 1.18  \\
              & 200      & -17$\pm$5$^{\ast}$ & -51.3$\pm$0.2$^{\ast}$ &  1.39     & 1.16   \\
              &  50              & -15$\pm$4$^{\ast}$    & -51.3$\pm$0.4$^{\ast}$ &  1.31     & 1.16    \\
              &  25              & -23$\pm$4             & -51.2$\pm$0.4$^{\ast}$ &  1.22     & 1.22     \\
              &  10              & -17$\pm$4$^{\ast}$    & -51.3$\pm$0.4$^{\ast}$ &  1.30     & 1.18      \\
              &  1              & -23$\pm$4             & -51.3$\pm$0.4$^{\ast}$ &  1.36     & 1.18       \\
\medskip      \\
260.3$\pm$0.6 &  $\infty$         & -24$\pm$3             & -49.1$\pm$0.2          &  1.30     & 1.35      \\
              &   200              & -26$\pm$3             & -49.1$\pm$0.2          &  1.28     & 1.36      \\
              &   50              & -24$\pm$3             & -49.1$\pm$0.2          &  1.33     & 1.35       \\
              &   25              & -24$\pm$3$^{\ast}$    & -49.1$\pm$0.3$^{\ast}$ &  1.29     & 1.35       \\
              &   10              & -26$\pm$3             & -49.1$\pm$0.2$^{\ast}$ &  1.31     & 1.35        \\
              &   1              & -23$\pm$3             & -49.1$\pm$0.2$^{\ast}$ &  1.34     & 1.34         \\
\medskip      \\
295.3$\pm$0.4 &    $\infty$         & 3$\pm$1               & -46.9$\pm$0.1     &    1.47   &            \\
              &    200              & 2$\pm$1               & -46.9$\pm$0.2     &    1.46   &             \\
              &    50              & 2$\pm$1               & -46.9$\pm$0.2     &    1.46   &             \\
              &    25              & 2$\pm$1               & -46.9$\pm$0.1     &    1.47   &             \\
              &    10              & -1$\pm$1              & -46.9$\pm$0.1     &    1.46   &             \\
              &    1              & 1$\pm$1               & -46.9$\pm$0.1     &    1.46   &             \\
\hline
\end{tabular}
\end{table}

\newpage

\begin{center}
{\bf Figure Captions}
\end{center}

\begin{enumerate}
\item Comparison of power spectra of the tagged particle potential energy 
fluctuations from the NVE ensemble and from NVT runs using
different values of the Berendsen coupling constant along the
1 g cm$^{-3}$ isochore at (a) 230K, (b) 260K and (c) 295K.
Insets show comparison over a relatively small frequency range 
between the NVE results and the results for a single value of $\tau_B$.
The figure key is the same in all cases.

\item Mean square displacement (MSD), in units of \AA$^2$,
 as a function of time (in ps) for different
values of the Berendsen coupling
constant at 230K and 1 g cm$^{-3}$. Inset shows the data using
the logarithmic scale.

\item Comparison of power spectra of the tagged particle potential energy 
fluctuations from the NVE ensemble and from NVT runs using
different values of the Berendsen coupling constant at
1 g cm$^{-3}$ and 230K. The power spectra for different values of $\tau_B$
are multiplied by different  factors to facilitate comparison
of the overall shapes of the curves.

\item Comparison of the power spectra of the total
 potential energy fluctuations at 295K
and 1 g/cm$^3$ for different values of the Berendsen coupling constant. 

\end{enumerate}
\vfill\eject
\newpage
\includegraphics{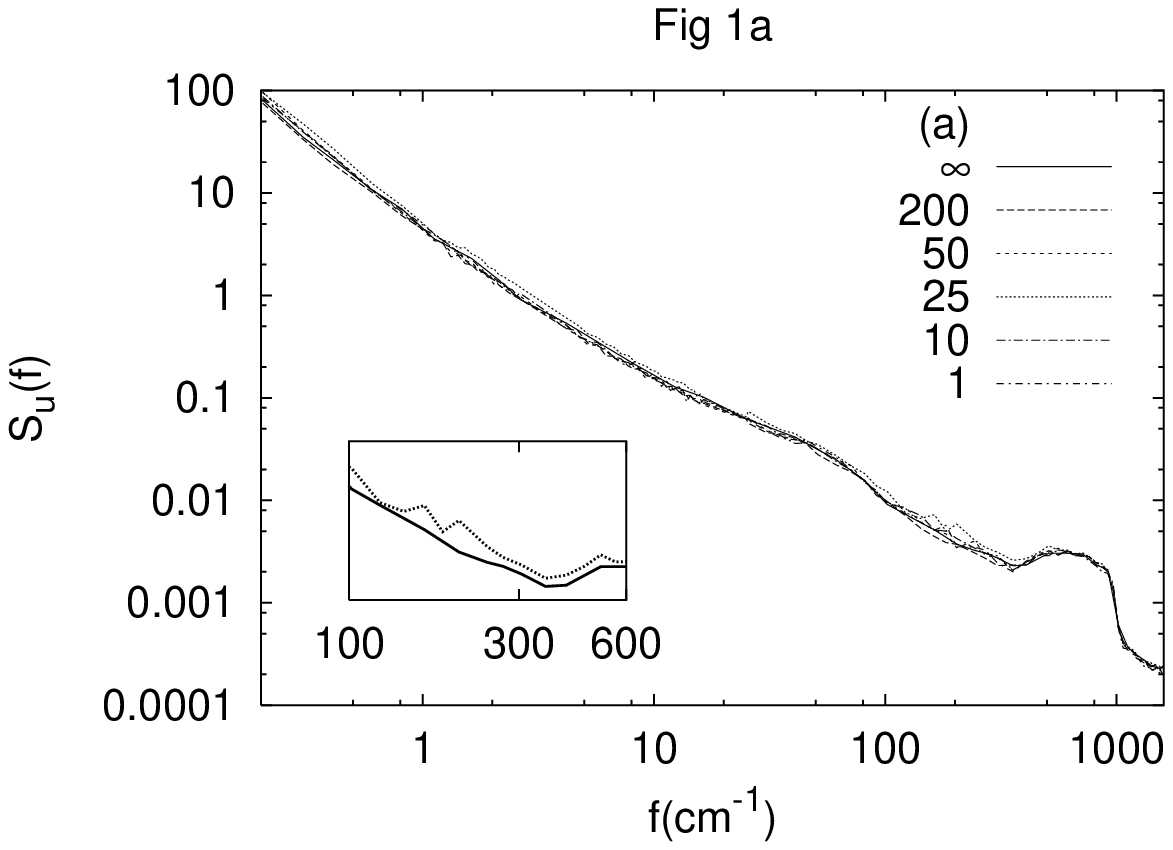}

\newpage
\includegraphics{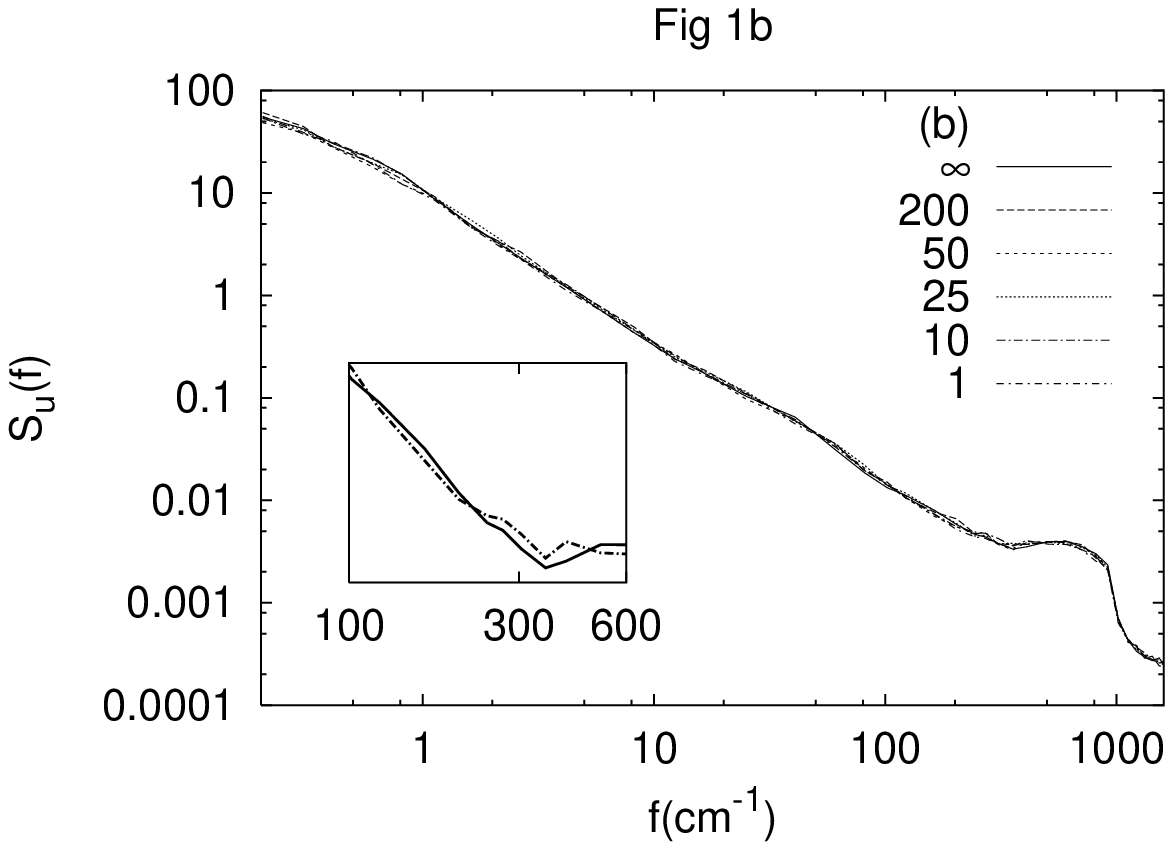}

\newpage
\includegraphics{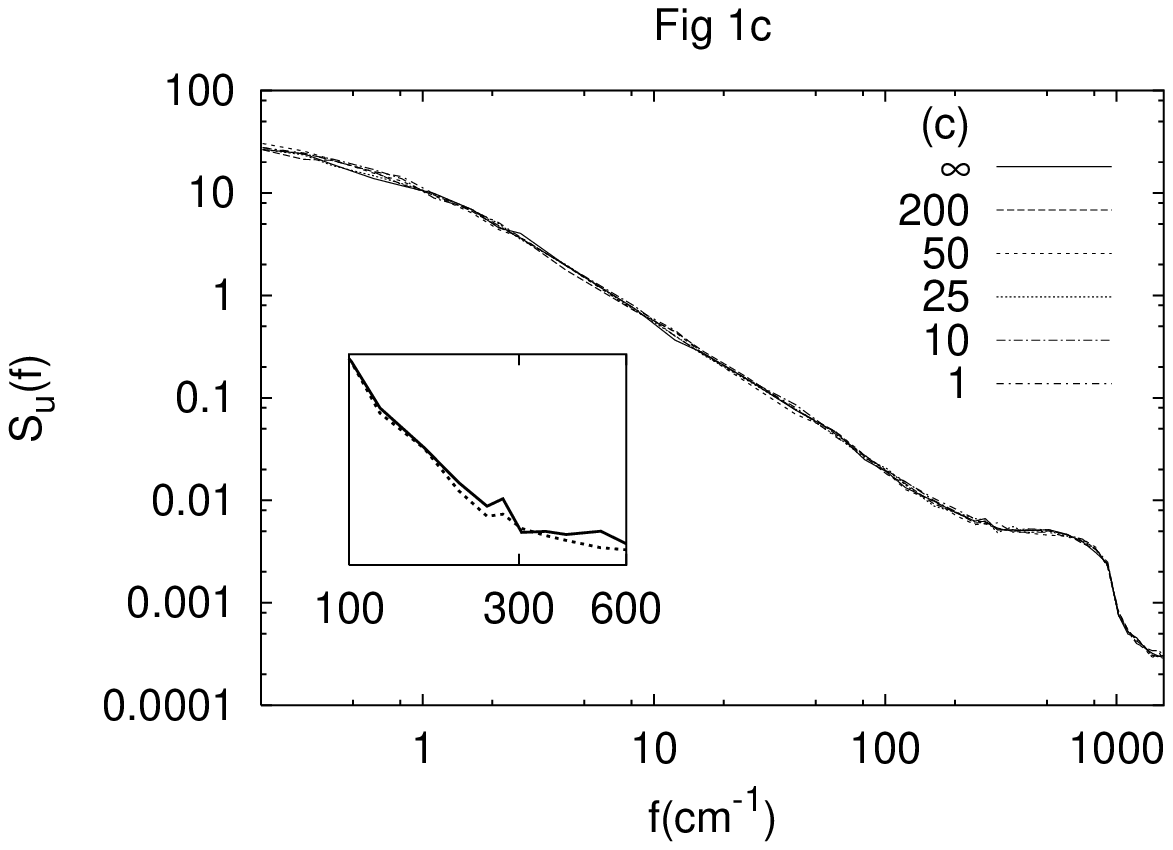}

\newpage
\includegraphics{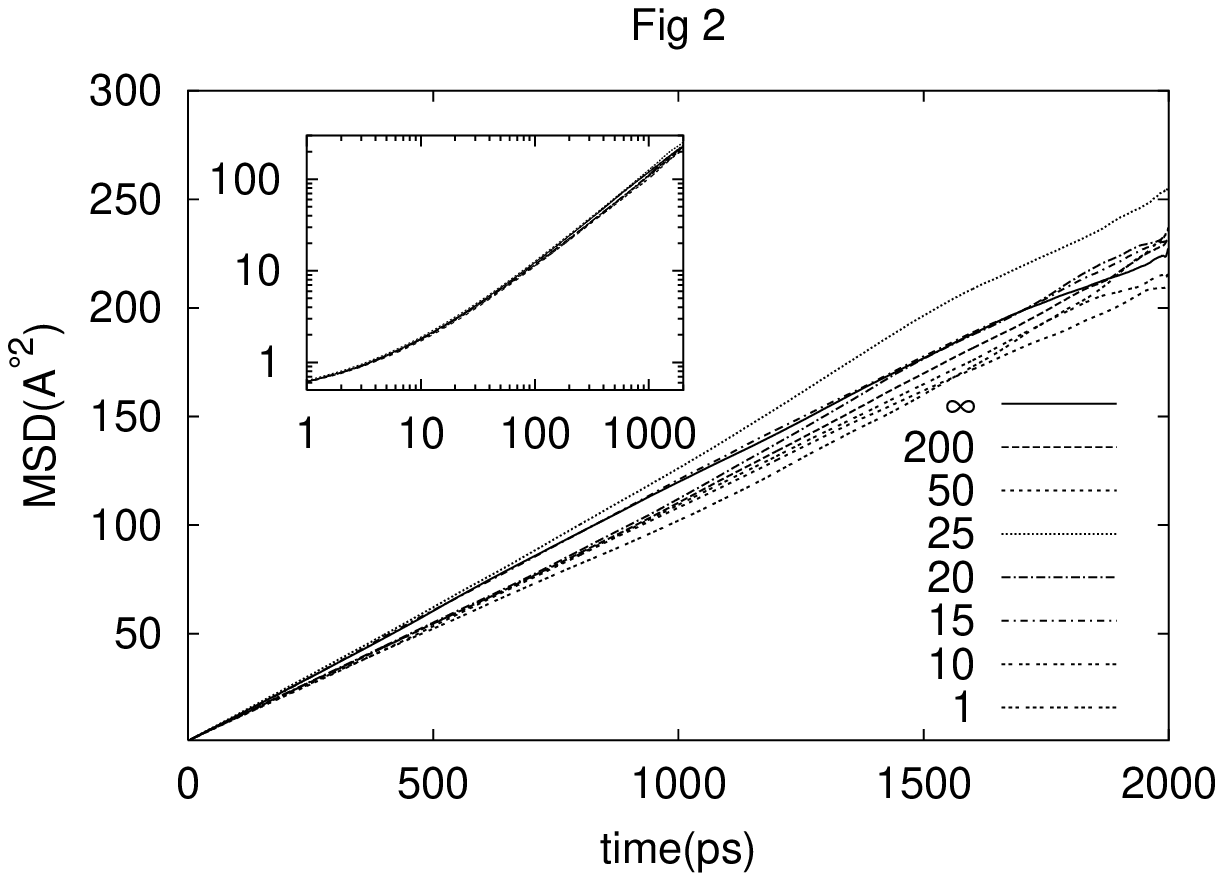}

\newpage
\includegraphics{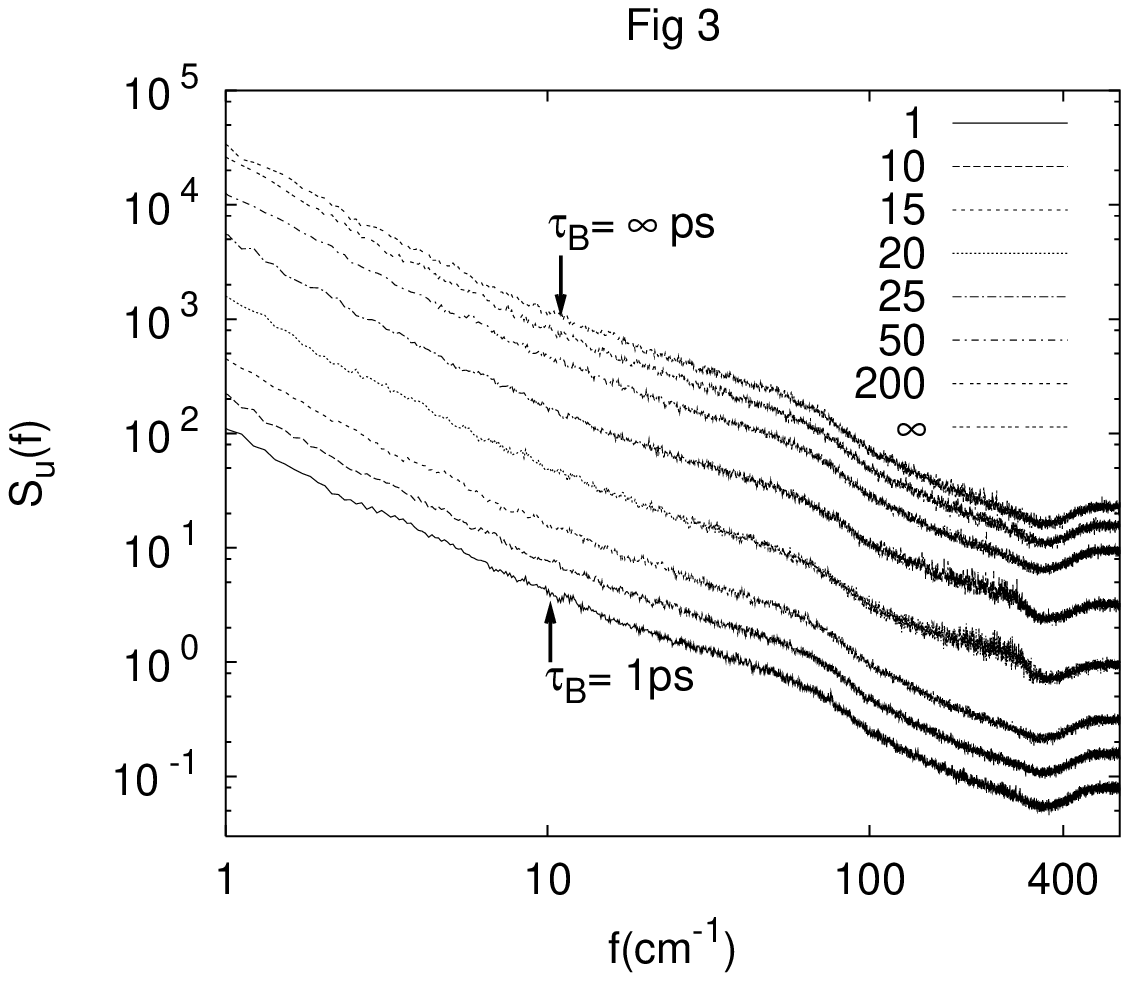}

\newpage
\includegraphics{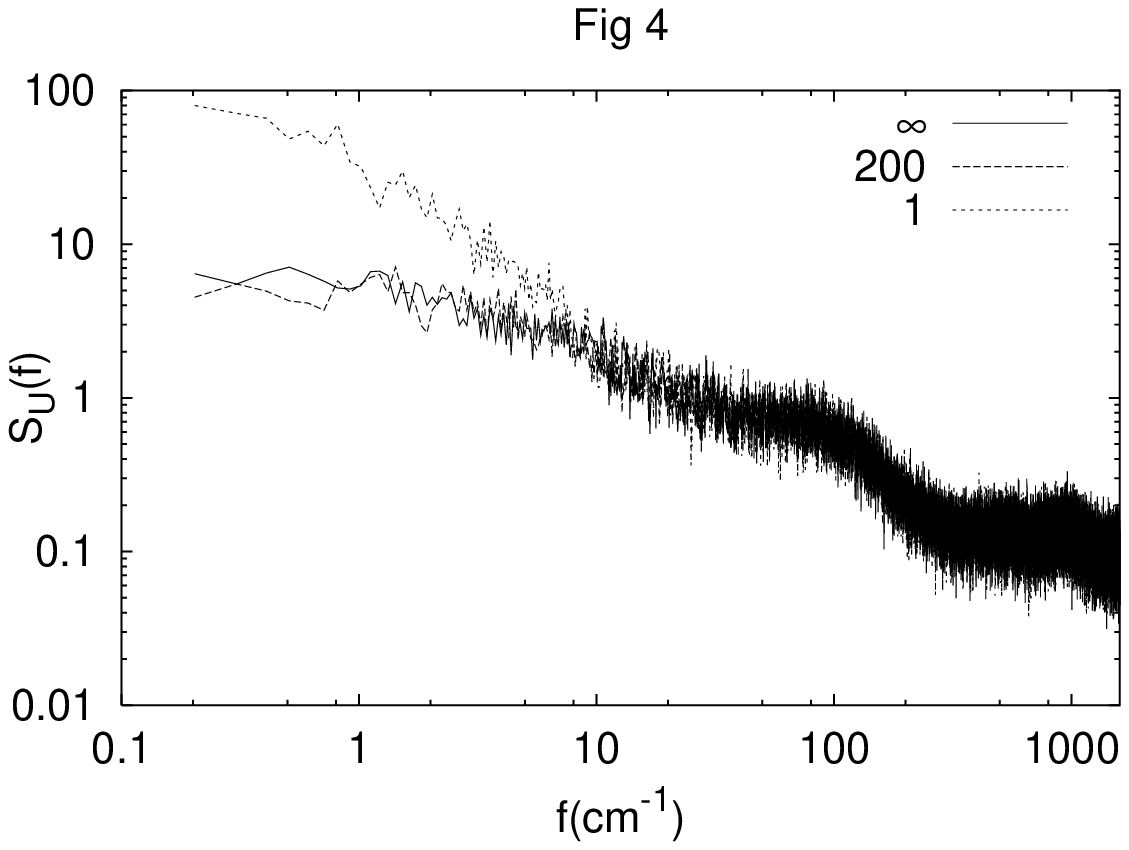}

\end{document}